# Control of surface potential at polar domain walls in a nonpolar oxide

G. F. Nataf[1,2,3,*], M. Guennou[2], J. Kreisel[2,4], P. Hicher[5], R. Haumont[5], O. Aktas[6,7], E.K.H. Salje[2,6], L. Tortech[8], C. Mathieu[1], D. Martinotti[1], N. Barrett[1]

[1]SPEC, CEA, CNRS, Université Paris-Saclay, CEA Saclay, 91191 Gif-sur-Yvette Cedex, France

[2]Materials Research and Technology Department, Luxembourg Institute of Science and Technology, 41 rue du Brill, L-4422 Belvaux, Luxembourg

[3]Department of Materials Science, University of Cambridge, 27 Charles Babbage Road, Cambridge CB3 0FS, UK

[4]Physics and Materials Science Research Unit, University of Luxembourg, 41 Rue du Brill, L-4422 Belvaux, Luxembourg

[5]Laboratoire de Physico-Chimie de l'Etat Solide, ICMMO, CNRS-UMR 8182, Bâtiment 410-Université Paris-Sud XI, 15 rue Georges Clémenceau, 91405 Orsay Cedex, France

[6]Department of Earth Sciences, University of Cambridge, Downing Street, Cambridge CB2 3EQ, UK

[7]State Key Laboratory for Mechanical Behavior of Materials & School of Materials Science and Engineering, Xi'an Jiaotong University, 710049 Xi'an, China

[8]IPCM, UMR CNRS 7201, UPMC, Université Pierre et Marie Curie, F-75005 Paris, France

*corresponding author: gn283@cam.ac.uk

**Abstract**

Ferroic domain walls could play an important role in microelectronics, given their nanometric size and often distinct functional properties. Until now, devices and device concepts were mostly based on mobile domain walls in ferromagnetic and ferroelectric materials. A less explored path is to make use of polar domain walls in nonpolar ferroelastic materials. Indeed, while the polar character of ferroelastic domain walls has been demonstrated, polarization control has been elusive. Here, we report evidence for the electrostatic signature of the domain-wall polarization in nonpolar calcium titanate ($CaTiO_3$). Macroscopic mechanical resonances excited by an ac electric field are observed as a signature of a piezoelectric response caused by polar walls. On the microscopic scale, the polarization in domain walls modifies the local surface potential of the sample. Through imaging of surface potential variations, we show that the potential at the domain wall can be controlled by electron injection. This could enable devices based on nondestructive information readout of surface potential.

# I. INTRODUCTION

Ultrahigh storage density combined with low power consumption is a major challenge for microelectronics in order to enable downscaling[1]. Domain-wall (DW) engineering[2] in ferroic materials is one possible route where the DW rather than the bulk material becomes the active element. DWs are regions where the changes of the order parameter from one ferroic domain to another result in strong gradient effects. The formation of DWs is a natural process by which a ferroic material minimizes its energy. Compared with domains, DWs are much narrower, inhomogeneous, and may have lower symmetry and completely different static or dynamic physical properties. The challenge is to observe, to predict, and to control the nanoscale DW functionality[3]. The electric conductivity of DWs in ferroelectrics is one functionality that is now well established and has been intensely studied[4–11]. Charged DWs can show conductivity many orders of magnitude higher than in bulk domains or in neutral DWs[8], providing a potential route towards nanometric metallic sheets in a dielectric matrix. For such materials, strategies to adjust the electronic conductivity[12,13] and proofs of concept for device applications have been reported[4,14,15].

Polarity of DWs in a nonpolar matrix has also been identified as an alternative route for DW electronics. The polar character of ferroelastic DWs has been proposed theoretically by using symmetry arguments[16] and their potential for memory devices has been discussed[17,18]. Despite this potential, they have been less studied experimentally. Most work has been devoted to the perovskite $CaTiO_3$ (CTO)[19,20]. It consists of corner linked $TiO_6$ octahedra with Ca atoms sitting in-between. It is distorted from the ideal cubic perovskite by two independent tilts of the octahedra network, and is described at ambient pressure and temperature in a *Pbnm* symmetry and an octahedral tilt system given by $a^-a^-c^+$ in Glazer notation[21] (Fig. 1(a,b)). The tilts are the primary order parameters that determine the distortion. As a mechanism for DW polarity, it has been proposed that one of the tilts goes to zero at the DWs, allowing for the emergence of Ti off-centering and hence DW polarization, seen as a competing secondary order parameter[22,23] (Fig. 1(c)). A biquadratic coupling between the primary order parameters and polarization is always allowed, and yields two energetically equivalent ground states for the wall polarity, plus and minus[24]. However, in the vicinity of DWs the main order parameter changes rapidly and may induce strong polarization via linear flexoelectricity[25]. Such gradient effects break inversion symmetry and therefore favor a specific polarization direction[24]. Whether or not a DW polarization can be switched therefore remains an open question.

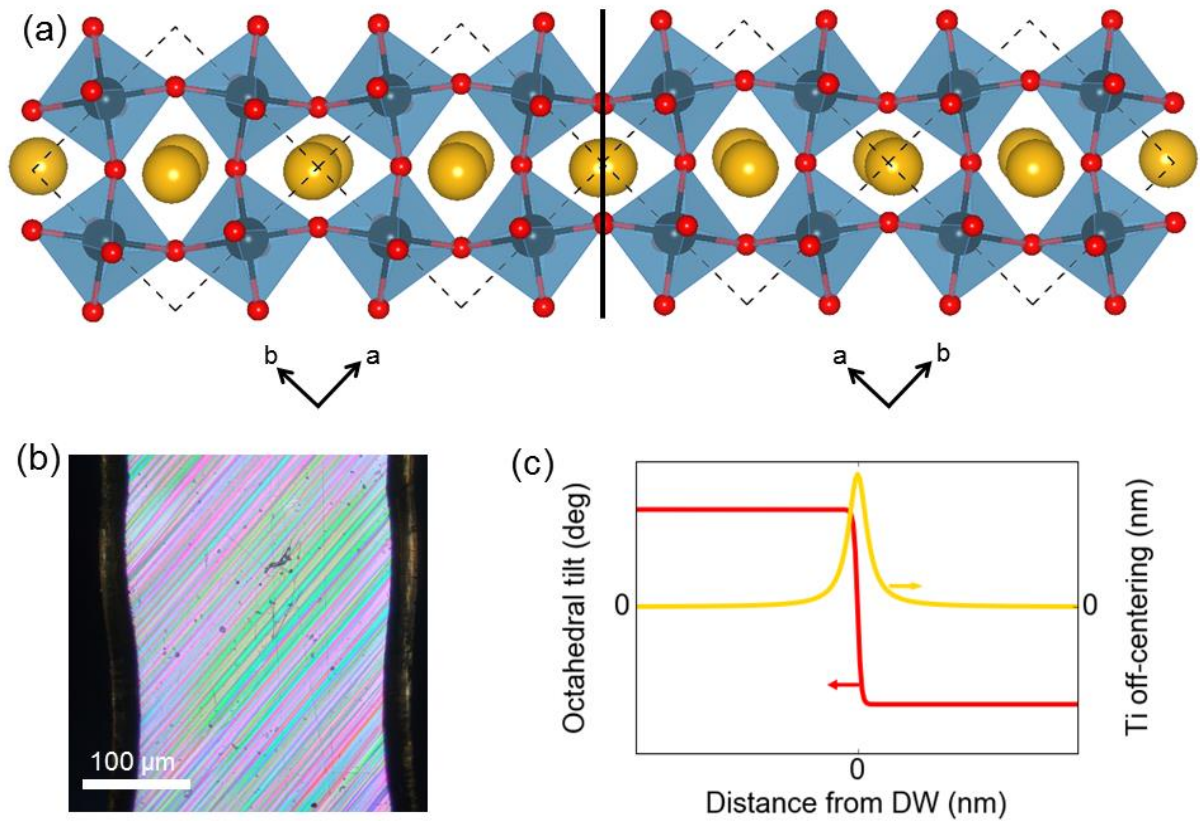

FIG. 1. Ferroelastic domain structure of CTO. (a) Stick-and-ball model of two ferroelastic domains in CaTiO$_3$, with the *Pbnm* cells mirrored with respect to the [1-10]$_{pc}$ plane. Large (yellow), medium (blue), and small (red) balls refer to Ca, Ti, and O ions, respectively. (b) Optical transmission image of a CaTiO$_3$ single crystal showing a complex ferroelastic domain structure. (c) Octahedral tilt (red) and Ti off-centering (yellow) as a function of the distance from the DW, illustrating schematically the emergence of cation displacement at a domain wall[16].

Aberration-corrected transmission electron microscopy has measured significant Ti off-centering at the DW in CTO[20]. Based on the atomic displacement parallel to the DW (about 6 pm), a polarization in the DW of 0.04 to 0.2 C m$^{-2}$ is expected, close to the bulk polarization in, for example, barium titanate. Second harmonic generation (SHG) provides another proof of the loss of inversion symmetry[19] but with a resolution of 0.5 μm along the lateral direction and 4 μm along the axial direction. For potential device applications, surface effects are also important. Eliseev *et al.* have carried out a theoretical study of the DW/surface junction in CTO[26] and suggest that rotostrictive and flexoelectric couplings lead to an electric field proportional to the structural deformation in the vicinity of the DW, providing a surface enhancement of the polarity. In spite of these advances, the control of the DW polarization has, so far, been elusive.

Imaging electric charge on a local scale is not trivial. Electron imaging of charged (ferroelectric) surfaces was proposed by Le Bihan and successfully applied to visualize domains in barium titanate, triglycine sulfate, and guanidine aluminum sulfate hexahydrate[27]. It relied on the use of a scanning electron microscope with strong interaction between the incident electrons and the insulating sample. A more recently introduced technique for studying ferroelectric and other polar materials is low-energy electron microscopy (LEEM). It provides full-field, noncontact imaging of surface potential with a spatial resolution typically better than 20 nm[28], almost 2 orders of magnitude better than, for example, SHG. At very low kinetic energies, also called the start voltage (SV), incident electrons are reflected before reaching the surface; this is mirror electron microscopy (MEM). At higher energies, electrons penetrate the sample surface and are elastically backscattered (LEEM). The transition between reflection and backscattering, the MEM-LEEM transition, provides a direct measure of the surface potential, which can be related to the surface polarization charge[29–31]. Ferroelastic DWs are expected to be polar, pointing either outwards or inwards toward the surface, giving rise to a net positive or negative charge at the surface. In addition, the difference in shear strain between adjacent domains imposes a shallow rooflike ridge/valley surface topography.

We use MEM and LEEM to study DW polarity at the surface of CTO. The intrinsic sensitivity of the technique to surface charge resolves outward- and inward-pointing polarizations. DWs are identified by symmetry, and their ridge/valley topography is visualized using atomic force microscopy (AFM). Low-energy electrons are then injected into the surface region, screening the DWs with positive surface charges. The original surface charge is fully recovered on annealing to 300 °C, showing that screening occurs through charge trapping, leaving the underlying DW polarity intact.

## II. EXPERIMENTAL

LEEM investigation of DWs was performed on a $CaTiO_3$ single crystal grown by the floating-zone technique with powders of $CaCO_3$ and $TiO_2$ as starting materials. The sample was polished with diamond paste and colloidal silica in order to reach a rms roughness of 0.45 nm. The sample was then annealed under oxygen at 1350 °C during 32 h. After this treatment, a complex ferroelastic domain structure was optically observed in the sample. The Laue photographs were recorded in backscattering geometry using a molybdenum x-ray source (0.4 Å < $\lambda$ < 2 Å). X-ray photoelectron spectroscopy was carried out using a monochromatic Al K$\alpha$ source (1486.7 eV) and an Argus 128 anode analyzer (ScientaOmicron). The base pressure was $2 \cdot 10^{-10}$ mbar. Spectra were recorded at room temperature.

Resonant piezoelectric spectroscopy is based on the excitation of elastic waves via piezoelectric coupling inherent to the sample. A small ac voltage (20 V) is applied across the sample, which is balanced between the ends of two piezoelectric transducers. The driving voltage leads to the excitation of local distortions that, when collective, lead to macroscopic resonant elastic waves. Any mechanical resonance is transmitted from the sample to the receiver transducer attached to the sample inside a He cryostat, similar to resonant ultrasound spectroscopy (RUS)[32,33].

The low-energy electron imaging experiments were performed using a LEEM III microscope (Elmitec GmbH), base pressure $1 \cdot 10^{-9}$ mbar. The incident electron beam was emitted by a thermionic $LaB_6$ electron gun at an accelerating voltage of $U_0 = -20$ kV. The sample was at $U_0 + SV$, where the Start Voltage (SV) defines the incident electron energy with respect to the sample surface. At low start voltages (SV < 0V), the incident electrons are reflected by the potential above the surface, *i.e.* Mirror Electron Microscopy (MEM). At higher voltages (typically, 1 V < SV < 10 V) they penetrate the sample and backscattering occurs, *i.e.* Low Energy Electron Microscopy (LEEM). The position of the MEM-LEEM transition measures the surface potential. The reflected or backscattered electrons are reaccelerated into the objective lens and finally pass through the imaging column. A double multichannel plate, screen, and camera recorded the electron intensity as a function of SV in a typical field of view of a few tens of microns. An angle-limiting aperture in the back focal plane of the objective lens cut off highly deviated electrons and improves spatial resolution. A schematic of the setup is shown in Ref. 28.

## III. RESULTS

### A. Macroscopic piezoelectric response of domain walls

Resonant piezoelectric spectroscopy (RPS) measures characteristic frequencies of mechanical resonances in a sample upon excitation by an ac electric field (Fig. 2(a)). Several resonances were observed in the CTO single crystal. We focus the analysis on the sharp peaks in the region around 870 kHz (Fig. 2(c)). They shift to lower frequencies and show amplitude variation with increasing temperature. We fitted the resonances with an asymmetric Lorentzian and extract the square frequency $f^2$ (proportional to the elastic moduli) and the inverse mechanical factor $Q^{-1}$ (proportional to the elastic losses). Both $f^2$ and $Q^{-1}$ are plotted as a function of temperature in Fig. 2(b). The squared frequency decreases almost linearly with increasing temperature. The elastic losses are low between 10 K and 100 K but show a rapid increase above 100 K. This is consistent with the behavior observed previously by resonant ultrasound spectroscopy in CTO[34] and pulse-echo ultrasonic technique[35].

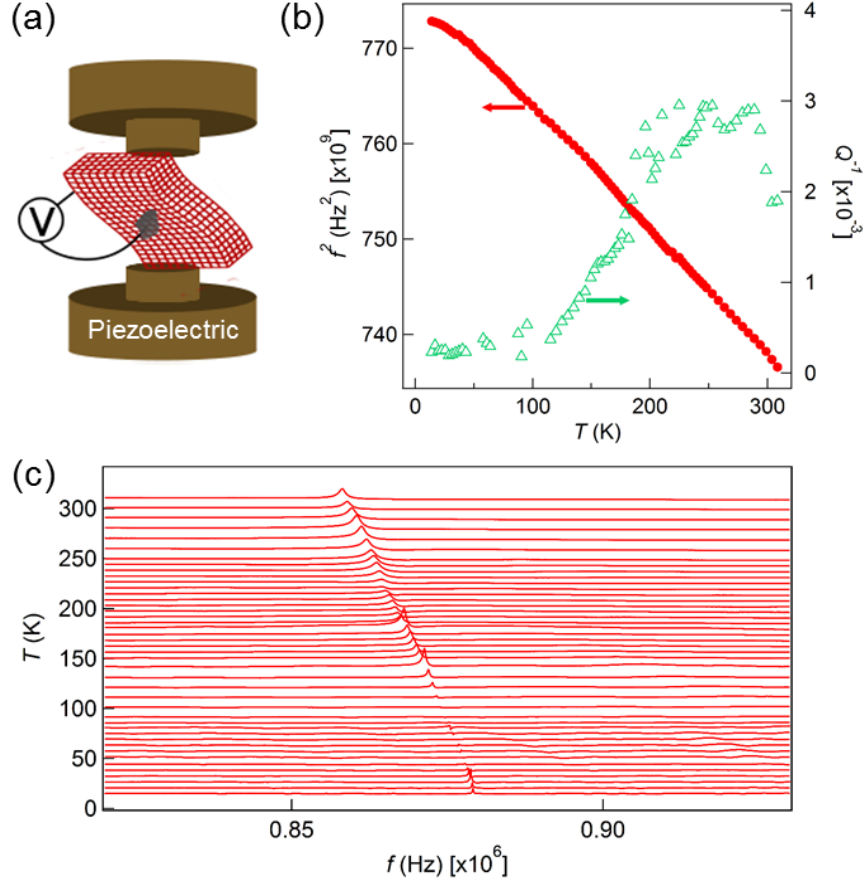

FIG. 2. Mechanical resonances induced by the piezoelectric effect. (a) Experimental setup for resonant piezoelectric spectroscopy. (b) Temperature dependence of a resonance near 870 kHz. Squared frequency is indicated by filled circles and the inverse mechanical factor by empty triangles. (c) Resonance amplitude as a function of frequency for temperatures between 10 and 310 K (as previously published in Ref. 36).

The RPS response with respect to the electric field is linear and we can therefore exclude electrostriction, which would have a quadratic dependence. Thus, the ac-field excitation of mechanical resonances provides evidence for a piezoelectric response within the material. We attribute this piezoelectric signature to the DWs, as previously described for $SrTiO_3$ (Ref. 37):

the applied ac field leads to the breathing of polar DWs through the piezoelectric effect, creating strain fields around them. The resulting elastic wave becomes resonant at a natural frequency, enhancing considerably the amplitude of the elastic wave. Although the fraction of DWs is small with respect to the bulk, it appears that the frequency range near 1 MHz, combined with a small imposed electric field (20 V), provides optimized sensitivity to microstructure dynamics, as observed, for example, in $SrTiO_3$ (Ref. 37).

Two other mechanisms could be invoked to explain the RPS response in a nonferroelectric phase: (i) polar defects as in the incipient ferroelectric (nonferroelastic) $KTaO_3$ and (ii) polar nanoregions as in the cubic phase of $BaTiO_3$ or of $Pb(Sc,Ta)O_3$[38,39]. Neither are probable. In the case of $KTaO_3$, above 120 K, the coherence of defect dipoles is low and the RPS signal weak[40]. With decreasing temperature, switchable defect dipoles freeze in parallel arrangements and induce macroscopic polarity, accompanied by an increase of the piezoelectricity. However, polar defects and polar nanoregions have not been reported in the literature on CTO. Therefore we conclude that the RPS resonances are indeed induced by polar, and hence piezoelectric, DWs. The observation of the piezoelectric response of the DWs is an experimental insight into the *dynamic* response of a DW under application of an electric field, as compared to the previous *static* observations[19,20].

It is interesting to note that the macroscopic piezoelectric response, and hence the RPS signal, should vanish if polar domain walls were statistically perfectly distributed. But this compensation is known to be usually imperfect in piezoelectric materials, and it is very likely that domain walls are distributed unequally, especially considering the comparatively large domain sizes, thereby explaining the existence of the RPS signal.

## B. Domain wall imaging by atomic force microscopy

The orientation, as determined by Laue diffraction, is along the [111] pseudo-cubic (pc) direction, with a 8° miscut. Fig. 3(a) is an optical image of the surface, different domains and DWs of various orientations are visible thanks to birefringence. The DW orientation of the DW with respect to the in-plane [1-10]$_{pc}$ direction taken as a reference is shown in Fig. 3(b). Four distinct angles are observed: 13°, 40°, 63° and 133°. The angular precision is ± 1°. Fig. 3(c) shows a three-dimensional (3D) landscape of the same area, acquired by AFM. The valleys (V) and ridges (R) of the twin structure appear clearly. The dots are surface contamination, which are also visible in the electron images and play no role in the observed DW properties.

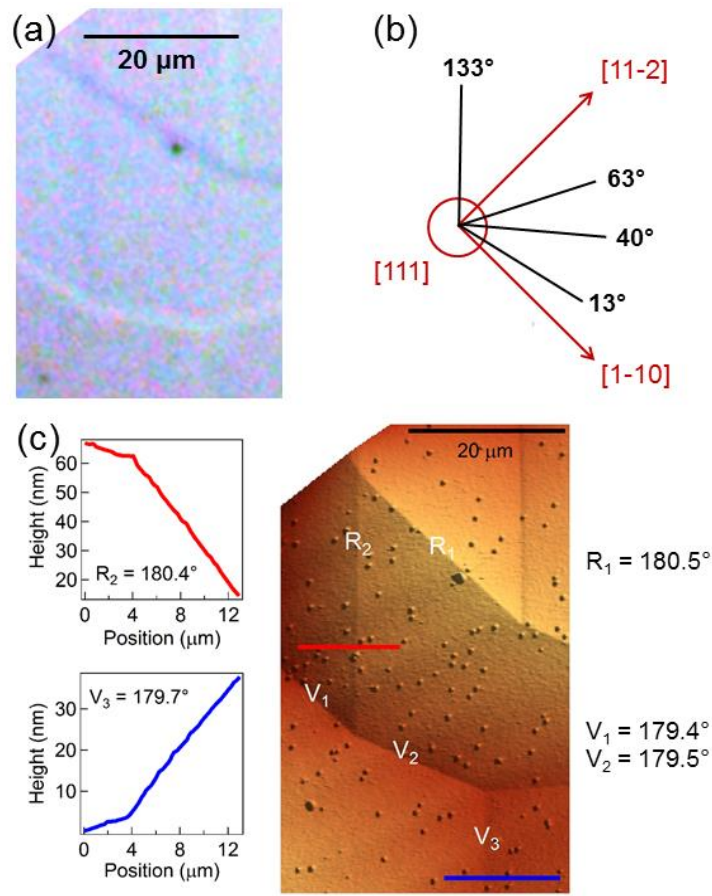

FIG. 3. Orientation and topography. (a) Optical micrograph of the surface (b) with DWs showing their angles with respect to [1-10]$_{pc}$ crystallographic orientation. (c) AFM topography of the sample surface. The valleys and ridges of the DWs are labelled "V" and "R", respectively. The upper inset shows the height profile perpendicular to the R$_2$ ridge, the lower inset the V$_3$ height profile.

Five DWs were selected and studied in more details by MEM and LEEM: ridges $R_1$ and $R_2$ and valleys $V_1$, $V_2$ and $V_3$. The insets show height profiles along the red ($R_2$) and blue ($V_3$) horizontal lines on the main image. The red (ridge) profile shows a 180.4° twin angle, whereas the blue (valley) profile is at 179.7°. Other angles ($R_1$, $V_1$ and $V_2$) are found in the range 179.4° – 180.5°.

## C. Expected domain wall orientation

We have calculated the possible orientations for the compatible ferroelastic DWs, following Ref. 41. The approach follows Yokota *et al.*[19], and requires that DWs are oriented to maintain strain compatibility between two adjacent domains, and therefore minimize stress and elastic energy. Based on these conditions, the equations for all "compatible" DWs can be derived. Two wall types are obtained: *W*, whose orientation is fixed by symmetry and *W'*, whose orientation depends on the coefficients of the spontaneous strain tensor describing the orthorhombic distortion. Once the equations of the DW planes are known, it is possible to calculate (i) the angle between the reference direction [1-10]$_{pc}$ and the DW as seen in microscopy or AFM, referred to as azimuthal angle, and (ii) the angle between the DW plane and the sample surface, referred to as inclination angle, and (iii) the angle of the ridges and valleys observed in AFM. The miscut is neglected.

For the calculation, we assume the same spontaneous strain tensor as considered in Ref. 19:

$$\begin{pmatrix} e_{11} & e_{12} & 0 \\ e_{12} & e_{11} & 0 \\ 0 & 0 & -2e_{11} \end{pmatrix},$$

where $e_{11} = 5.7 \cdot 10^{-4}$ and $e_{12} = 5.5 \cdot 10^{-3}$ as determined from the lattice parameters[42] according to expressions given in Ref. 43.

We enumerate the different possible domain pairs ($D_a$, $D_b$), and calculate for each pair the difference in their spontaneous strain tensors ($\Delta\varepsilon = \varepsilon_a - \varepsilon_b$). For each pair, the equations of the two possible compatibles DWs are derived classically by writing $\Delta\varepsilon_{ij}x_ix_j = 0$. From the equation of the DW plane, or equivalently a normal unit vector written $\mathbf{u}_{DW}$, and the equation of the plane of the sample surface (here $[111]_{pc}$ with normal unit vector $\mathbf{u}_{111}$), the azimuthal and inclination angles are calculated respectively as

$$\cos^{-1}\left(\frac{\boldsymbol{u}_{1-10} \cdot (\boldsymbol{u}_{111} \times \boldsymbol{u}_{DW})}{\|\boldsymbol{u}_{111} \times \boldsymbol{u}_{DW}\|}\right)$$

and

$$\cos^{-1}(\boldsymbol{u}_{111} \cdot \boldsymbol{u}_{DW})$$

where $\mathbf{u}_{1-10}$ here denotes a unit vector parallel to [1-10], which is the direction chosen as a reference in the $[111]_{pc}$ plane.

For the angles of the ridges and valleys, for each DW, we want to determine the angle formed in the orthorhombic phase by two unit vectors in $D_a$ and $D_b$ originally parallel in the cubic phase and chosen in such a way that: (i) they are contained in the $[111]_{pc}$ plane in the cubic phase (ii) they are orthogonal to the line formed by the intersection of the DW and the sample surface – since this is how the angle is measured in AFM. If the differential spontaneous strain tensor $\Delta\varepsilon$ is conveniently expressed in suitable axes ($\mathbf{u}_1,\mathbf{u}_2,\mathbf{u}_3$), with $\mathbf{u}_1$ parallel to $[111]_{pc}$, $\mathbf{u}_2$ aligned with the line formed by the DW on the $[111]_{pc}$ surface and $\mathbf{u}_3 = \mathbf{u}_1 \times \mathbf{u}_2$, this angle is simply given by $2\Delta\varepsilon_{13}$.

However, whether or not this angle corresponds to the angle observed in AFM depends on the assumption that the actual sample surface at the location of the DW in the cubic phase is perfectly flat. This is impossible to check experimentally, and most likely wrong, since the polishing of the crystal was done at room temperature and not in the cubic phase. Also, for simplicity, we considered in this calculation only nonordered domain pairs, i.e., ($D_a,D_b$) was not distinguished from ($D_b,D_a$), which results in an ambiguity on the signs of the angles.

Based on these results, we identify 6 possible DW orientations that correspond to the optical and AFM images, recognized by the values of their azimuthal angles (by symmetry, supplementary angles are equivalent). The inclination angles cannot be checked experimentally here, but it is interesting to note that both "straight" $W$ walls normal to the surface as well as inclined $W'$ walls are present in our area. The ridge and valley angles are difficult to assign conclusively, but are in the correct range.

| Equation | Azimuthal angle (°) | Inclination angle (°) | Ridge/valley angle (°) | Name |
|---|---|---|---|---|
| | | W-wall | | |
| x = 0 | 120 | 54.7 | 180.3/179.7 | |
| y = 0 | 120 | 54.7 | 179.7/179.7 | |
| z = 0 | 0 | 54.7 | 179.7/180.3 | |
| x = y | 90 | 90 | 180.3/180.7 | |
| x = -y | 0 | 35.3 | 180.3/179.7 | |
| **y = z** | **150** | **90** | **179.7/180.3** | **$V_2$** |
| y = -z | 60 | 35.3 | 179.7/180.3 | |
| **z = x** | **150** | **90** | **179.3/180.4** | **$V_2$** |
| z = -x | 60 | 35.3 | 179.7/180.3 | |
| | | W'-wall | | |
| $3e_{11}(z+x)+e_{12}y = 0$ | 120 | 42.3 | 180.0 | |
| $3e_{11}(z+x)-e_{12}y = 0$ | 60 | 112.9 | 180.0 | |
| $3e_{11}(z-x)+e_{12}y = 0$ | 104.9 | 55.7 | 180.3 | |
| **$3e_{11}(z-x)-e_{12}y = 0$** | **44.9** | **124.3** | **180.3** | **$R_2$, $V_3$** |
| $3e_{11}(x+y)+e_{12}z = 0$ | 0 | 42.3 | 180.0 | |
| $3e_{11}(x+y)-e_{12}z = 0$ | 0 | 112.9 | 180.0 | |
| **$3e_{11}(x-y)-e_{12}z = 0$** | **164.9** | **124.3** | **179.7** | **$R_1$, $V_1$** |
| **$3e_{11}(x-y)+e_{12}z = 0$** | **15.1** | **55.7** | **179.7** | **$R_1$, $V_1$** |
| $3e_{11}(y+z)+e_{12}x = 0$ | 120 | 42.3 | 180.0 | |
| $3e_{11}(y+z)-e_{12}x = 0$ | 60 | 67.1 | 180.0 | |
| **$3e_{11}(y-z)+e_{12}x = 0$** | **135.1** | **55.7** | **179.7** | **$R_2$, $V_3$** |
| $3e_{11}(y-z)-e_{12}x = 0$ | 75.1 | 124.3 | 179.7 | |

TABLE I. DW equations and characteristic angles. Azimuthal and inclination angles with respect to $[1\text{-}10]_{pc}$ direction and the (111) surface, and ridge/valley angles obtained assuming of a flat (111) surface in the cubic phase. For $W$ walls, two values are given corresponding to 2 different domain pairs. The bold characters indicate the walls identified in our sample.

The measured angles of $R_1$, $R_2$, $V_1$, $V_2$ and $V_3$ with respect to $[1\text{-}10]_{pc}$ are 13°, 35° and 133°. The closest azimuthal angles calculated in Table I are 15.1°, 30° (or 150°) and 135.1°, respectively, in good agreement given the approximations made in the calculations. The differences between measured and calculated angles may be due to the 8° miscut. Following the symmetry calculations of the *W* and *W'* twins we expect inclination angles of 55.7°, 90° and 124.3°, respectively.

**D. Electron imaging of domain walls**

Fig. 4 shows two MEM images taken at SVs of (a) −0.8 and (b) +0.3 V of the same area as the AFM landscape and optical micrograph. In the MEM images the electron intensity from the domain surfaces far from the DWs always has the same value, indicating identical surface potential. The DWs $R_2$-$V_1$-$V_2$-$V_3$ appear as dark lines whereas $R_1$ is bright. The white dots correspond to the surface contamination in the form of nanoparticles observed by AFM and provide an intrinsic benchmark to distinguish contrast due to physical and electrical topography.

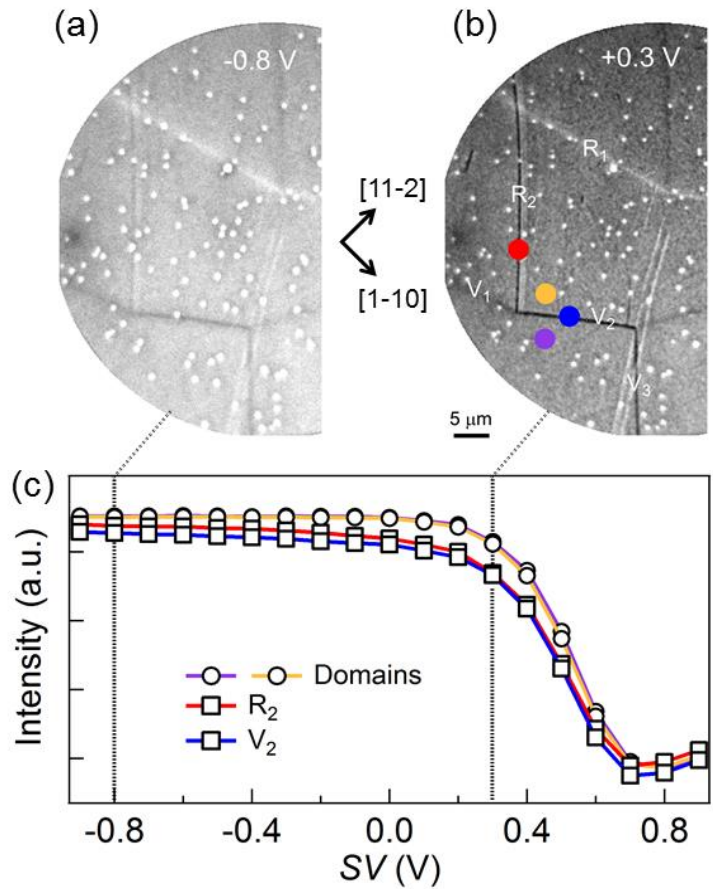

FIG. 4. MEM images of the DWs. SV of (a) −0.8 and (b) +0.3 V. (C) Electron intensity as a function of SV measured in domains and at DWs showing a 100 mV shift in the MEM-LEEM transition to lower SV for the DWs.

At a SV of -0.8 V electrons are almost completely reflected and the contrast between the DWs and domains is low (Fig. 4(a)). With increasing SV, electrons approach the surface and are more sensitive to local variations in the surface potential and the DW contrast increases as shown in Fig. 4(b) at +0.3 V. Fig. 4(c) plots the electron intensity in the domains and at the DWs as a function of SV. The MEM-LEEM transition occurs at a SV = 0.43 V in both domains showing that they have the same surface potential. At the $R_2$-$V_2$ DWs the MEM-LEEM transition occurs at lower (by about 100 mV) SV which is a signature of more positive surface charge at the DW with respect to the domains.

We have used slight under-focusing in order to enhance the contrast due to surface charge. As shown in Supplementary[44] Fig. S2 and S3, this identifies the dark DWs ($R_2$, $V_1$, $V_2$ and $V_3$) as having outwards pointing, *i.e.* positive, polarity, whereas $R_1$ has an inward pointing polarity, *i.e.* negative surface charge. Thus, both ridge and valley DWs may adopt the same polarity, in agreement with theoretical calculations of the wall energies[45]. Even more fascinating is that a ridge may have positive or negative polarity. The contrast is inverted for over-focusing (Fig. S2) which further confirms the DW polarity.

### E. Electron injection and screening of domain wall polarity

We now consider the influence of low energy electron injection on DW contrast in the electron images. We define the contrast at DWs as $C = (I_{\text{domain wall}}-I_{\text{domain}})/I_{\text{domain}}$, where $I_{\text{domain wall}}$ and $I_{\text{domain}}$ are the intensities at DWs and in domains, respectively. The magnitude, $|C|$, is different for each DW and is higher at $R_2$, $V_2$ and $V_3$.

Electron injection is accomplished by increasing the SV well beyond the MEM-LEEM transition so that the majority of incident electrons penetrate the sample surface. In Fig. 5(b,c) we show MEM images acquired at SV = 0.3 V following exposure to electrons with SV = 8 V for 2 and 32 minutes, respectively. After 2 minutes the contrast magnitude at valleys and at $R_2$ has diminished but has not changed at $R_1$. After 32 minutes of irradiation, the contrast at $R_2$ is zero, at valleys it is weaker but $R_1$ remains unchanged. Fig. 5(d) plots the quantitative evolution of the contrast with irradiation time for all of the DWs. It shows the quite distinct responses of $R_1$ and $R_2$ to electron injection, the contrast at $R_1$ is constant whereas that at $R_2$ decreases to zero. The contrast at valleys ($V_1$, $V_2$ and $V_3$) is attenuated by about 20% after 2 minutes exposure to the electron beam but then remains constant for longer exposure times.

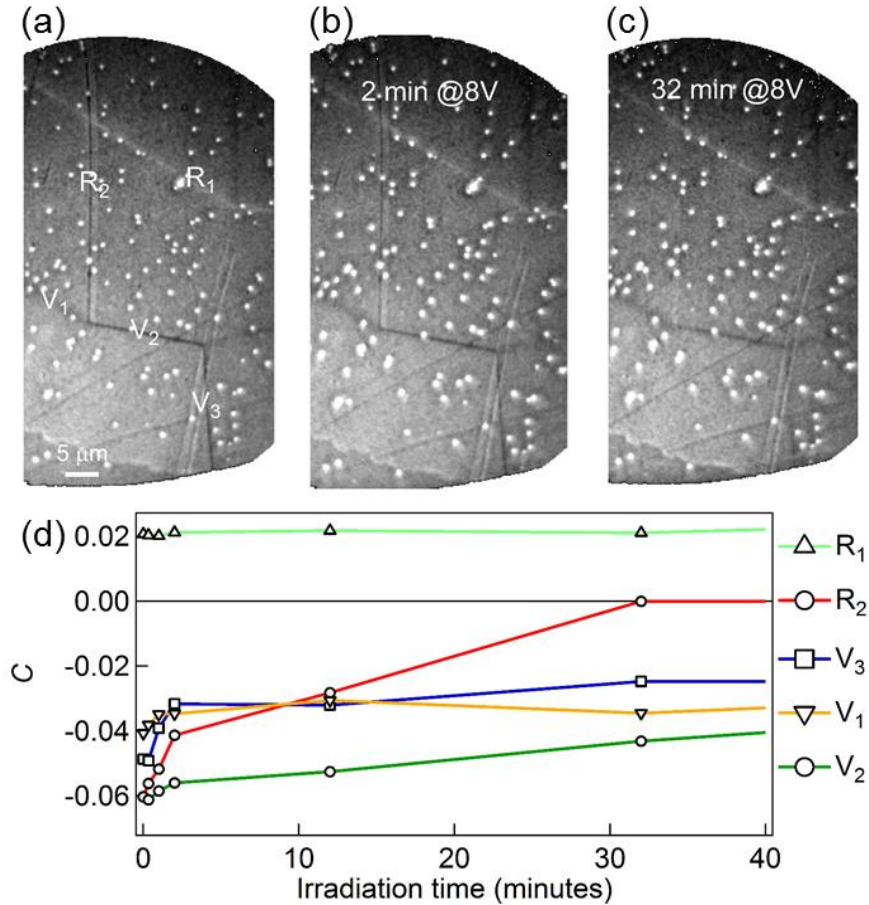

FIG. 5. Time-dependent MEM images. Images taken at SV = 0.3 V after exposure to 8 V electrons for (a) 0 (b) 2 and (c) 32 minutes. (D) Intensity contrast at DWs as a function of irradiation time.

We can estimate the total number of electrons injected in the sample through the beam spot as the difference between the incident and detected current, assuming the maximum reflectivity to be unity. For an electron gun emission current of 20 nA, after 2 minutes at 8 V, $7 \cdot 10^{12}$ electrons have been injected, increasing to $1.1 \cdot 10^{14}$ electrons after 32 minutes. However, electrons can be elastically or inelastically scattered and in the absence of an accurate drain current measurement these are only approximate values.

Following Cazaux[46] and Ziaja[47], the inelastic mean free path of the electrons at 4 V is 5-6 nm. Thus for a circular beam spot of diameter 90 μm, the injected charge density after 32 minutes at 8 V is $5.8 \cdot 10^{29}$ e cm$^{-3}$. This is 7 orders of magnitude higher than, for example, the injected charges used to switch ferroelectric and ferroelastic domains in BaTiO$_3$ in a transmission electron microscope[48], however, the energy of the injected electrons is five orders of magnitude smaller. The difference in electron energy and probable difference in radiation damage and secondary electron cascade make it difficult to further compare these two experiments.

### F. Reversibility of polarity screening

It is possible to recover the initial state of DW contrast before charge injection by annealing the sample above 80 °C. Figure 6 compares images taken after 10 min exposure to 20 V electrons (a–c) and images of the same regions after annealing at 330°C (d–f). Before annealing, the contrast at DWs with outward-pointing polarity is weak; only the DW ridge R$_1$ with inward-pointing polarity is clearly distinguishable. After annealing, the contrast at R$_2$, V$_1$, V$_2$ and V$_3$ has been fully recovered, while the contrast at R$_1$ is still present. Fig. 6(g) provides a quantitative view of the temperature dependence of the DW contrast for R$_1$, R$_2$, and V$_3$. While there is little change in the contrast of R$_2$ and V$_3$ up to 80°C, a clear evolution starts above 80°C with full contrast recovery at 330°C. The contrast then remains constant as the sample is cooled back down to room temperature. This is strong evidence that screening of the DW contrast is achieved by space-charge formation of injected electrons in the near-surface region. Once the space charge is set up, dissipation must be thermally activated. However, annealing up to 330°C only activates charge dissipation, leaving the DW position and polarity unchanged. Hence, there is a memory effect of the DW topography and polarity which are robust up to at least 330°C.

The recovery of contrast on annealing indicates that the space charge around positive-polarity DWs is dissipated by thermal activation, similar to the thermally stimulated currents measured in the nominally centric $(Ba,Sr)TiO_3$, one component of which was ascribed to the built-in polarization whereas the other was shown to be due to trapped charge[49].

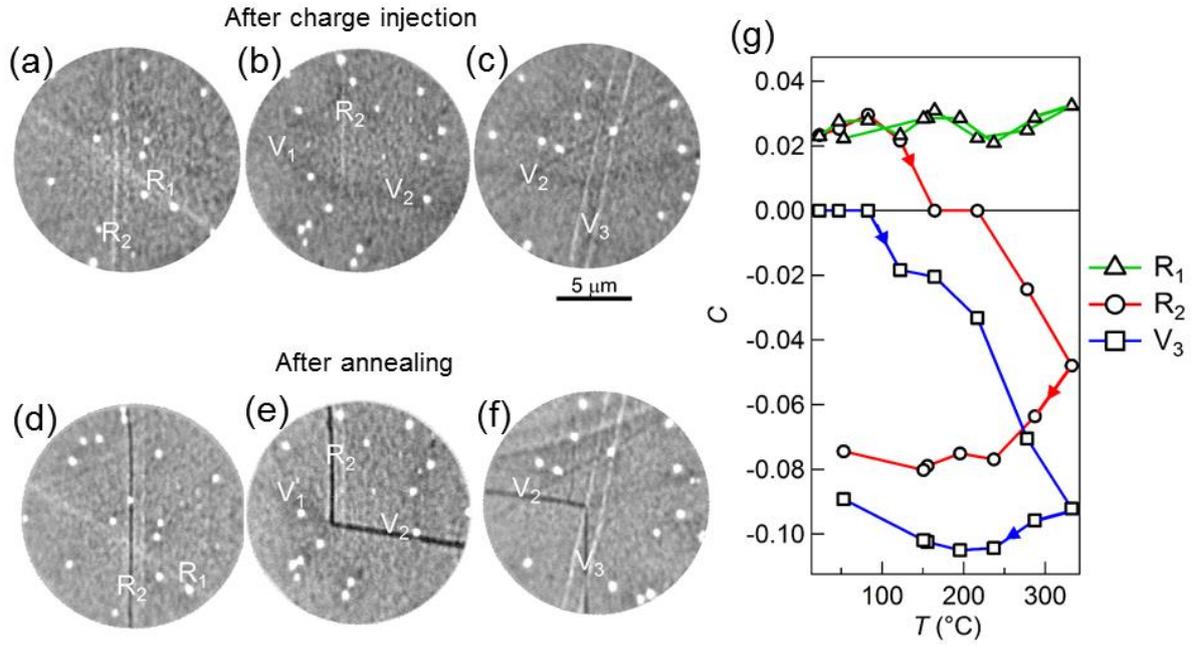

FIG. 6. Influence of annealing. MEM images acquired at a SV = 0.3 V after 10 minutes exposure to 20 V e-beam. (a) $R_1$-$R_2$ (b) $R_2$-$V_1$-$V_2$ and (c) $V_2$-$V_3$. (d-f) same regions after annealing at 330°C. (g) Intensity contrast at $R_1$, $R_2$ and $V_3$ as a function of temperature during annealing up to 330°C (right pointing arrow) and cooling back down to room temperature (left pointing arrow).

## IV. DISCUSSION

### A. Nature and direction of DW polarization

Let us first state how our observations provide evidence for DW polarization. Surface topography can be expected to play a role in MEM-LEEM imaging, but it cannot be the dominant origin for the contrast observed here. As shown in Fig. 5a, ridges $R_1$ and $R_2$ have opposite contrast with respect to surrounding domains. If the contrast was dominated by the physical topography, i.e., by the ridge, it should have the same sign. Since this is not the case, the ridge angle alone cannot account for the contrast. Instead, it is due to the surface charge induced by the DW polarization. Thus we show that both inward- and outward-pointing polarities are possible in both ridges and valleys, as predicted by theory in the case of $SrTiO_3$[45] If this is valid for CTO, then biquadratic coupling between the primary and secondary order parameters rather than gradient effects[24] makes the main contribution to the wall polarity.

The observation of charge variations associated with DW polarization raises the question of a possible ferroelectric, and not only polar, character of the DWs. The fact that we observe several DWs (*W* and *W'* types, ridge and valleys) and that they exhibit differences in their MEM contrast suggests that the polarization in the DW can be switched. However, uncertainties in DW assignment do not allow us to clearly identify the archetypal situation of opposite polarization signs in otherwise *strictly identical* DWs, i.e., separating identical domains with the same orientation in space, or conversely, identical polarization charges in opposite DWs. Besides, definitive evidence for ferroelectricity, i.e., polarization switching under electric field, has yet to be demonstrated. Further investigations are therefore required to conclusively establish the ferroelectric nature of DWs in CTO.

We can relate our observations to the atomistic picture drawn from aberration-corrected transmission electron microscopy (HR-TEM) measurements on a (110) DW[20]. Two systematic movements of the Ti atoms have been reported: a displacement perpendicular to the DW in the second closest layers pointing towards the DW, and a larger displacement parallel to the DW in the layers adjacent to the DW[20]. The former would give rise to small positive polarization charges on the sample surface independent of the DW orientation with respect to the surface. On the other hand, the much bigger displacement parallel to the DW is responsible for DW polarization and gives rise to surface charges that vary with the DW angle with respect to the sample surface. In the case of a (110) DW, HR-TEM imaging reveals that the polarization is along the $[1\text{-}10]_{pc}$ direction, as expected by symmetry. In our sample, this applies to the DW labelled $V_2$, and the projection of the DW polarization on the direction normal to the (111) surface plane would indeed be non-zero and give rise to contrast observable in LEEM, as shown in Fig. 7a and 7b. The observations of Van Aert *et al*. are therefore one of the cases of DW orientation with respect to the surface.

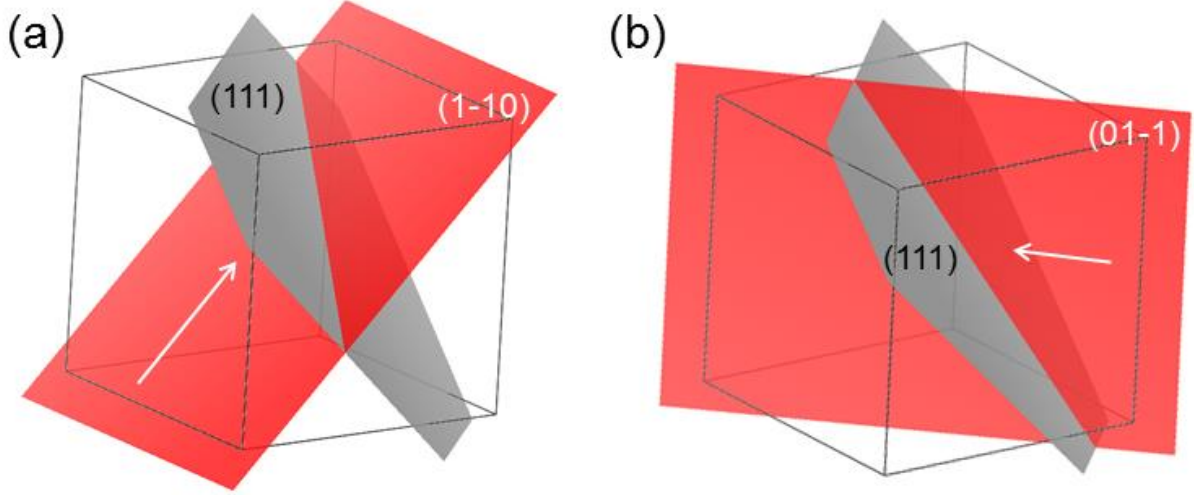

FIG. 7. Polarization charges at the surface (111). Schematics of the polarization in the DW $V_2$, following the assignation of Table I, (a) $z = x$ plane, (b) $y = z$ plane. The arrows indicate the polarization direction.

### B. Screening mechanism

The positive charge of outward-pointing polarity DWs is screened by electron injection. We suggest that electrons form a space-charge region near the junction between the outward-pointing polarity DWs and the surface. They screen the positive surface charge but not the negative charge at the junction of an inward-pointing polarity DW and the surface. Interestingly, the positive-polarity ridge is better screened than positive-polarity valleys. This may be due to local geometry. Near-neighbor surface oxygen displacements around the wall, for example, could impede establishment of a sufficiently dense space-charge region to fully screen the valley DWs at the surface.

It is known that oxygen vacancy ($V_O$) formation is favored at ferroelastic DWs[50]. There is no spectroscopic evidence for oxygen vacancies (Fig. S1) averaged over the whole sample, however, there may be a higher concentration of oxygen vacancies near the DWs. High spatial resolution chemical analysis would be necessary to determine this. In addition, vacancy migration towards the DWs may also be dependent on the wall inclination angle, providing a further possible explanation for the quantitative differences observed in positive polarity DW screening. This must be more systematically investigated in the future.

## V. CONCLUSION

Variations of potential at oxide surfaces can be used as a non-contact and non-destructive information readout method[51]. Here, we demonstrate that polar DWs locally modify the surface potential on a lateral scale well beyond the typical DW width of a few unit cells, providing an immediate and reliable readout of DW polarity. We also show that injecting electrons can be used to control the surface potential near positive-polarity walls. Two DWs with the same physical topography ($R_1$, $R_2$) can have opposite polarity and are stable. $R_1$, $R_2$ are both type $W'$ walls, suggesting it should be possible to switch the polarity in the same DW. $R_1$ and $R_2$ intersect, showing it may be possible to store opposite bits of information side-to-side. We believe that the feasibility of nonvolatile memories based on the surface potential and the characteristics of the ferroelastic DWs (potentially controllable position and inclination angle) may couple for new potential device applications. Local DW chemistry, in particular, oxygen vacancy concentration, may provide an additional handle in manipulating the surface potential.

## ACKNOWLEDGEMENTS

GFN and MG would like to thank the SPMS for giving access to their X-ray facilities. GFN, NB, MG and JK acknowledge support from the Luxembourg National Research Fund under project CO-FERMAT FNR/P12/4853155/Kreisel, and the ANR HREELM project. EKHS is grateful to the Luxembourg National Research Fund for support during his stay at LIST under project INTER/MOBILITY/13/6572599.

# REFERENCES

[1] S.P. Mohanty and A. Srivastava, *Nano-CMOS and Post-CMOS Electronics: Devices and Modelling* (2015).

[2] E.K.H. Salje, ChemPhysChem **11**, 940 (2010).

[3] G. Catalan, J. Seidel, R. Ramesh, and J.F. Scott, Rev. Mod. Phys. **84**, 119 (2012).

[4] J. Seidel, L.W. Martin, Q. He, Q. Zhan, Y.-H. Chu, A. Rother, M.E. Hawkridge, P. Maksymovych, P. Yu, M. Gajek, N. Balke, S. V Kalinin, S. Gemming, F. Wang, G. Catalan, J.F. Scott, N.A. Spaldin, J. Orenstein, and R. Ramesh, Nat. Mater. **8**, 229 (2009).

[5] J. Guyonnet, I. Gaponenko, S. Gariglio, and P. Paruch, Adv. Mater. **23**, 5377 (2011).

[6] S. Farokhipoor and B. Noheda, Phys. Rev. Lett. **107**, 127601 (2011).

[7] J. Seidel, P. Maksymovych, Y. Batra, A. Katan, S.-Y. Yang, Q. He, A.P. Baddorf, S. V. Kalinin, C.-H. Yang, J.-C. Yang, Y.-H. Chu, E.K.H. Salje, H. Wormeester, M. Salmeron, and R. Ramesh, Phys. Rev. Lett. **105**, 197603 (2010).

[8] T. Sluka, A.K. Tagantsev, P. Bednyakov, and N. Setter, Nat. Commun. **4**, 1808 (2013).

[9] M. Schröder, A. Haußmann, A. Thiessen, E. Soergel, T. Woike, and L.M. Eng, Adv. Funct. Mater. **22**, 3936 (2012).

[10] D. Meier, J. Seidel, A. Cano, K. Delaney, Y. Kumagai, M. Mostovoy, N.A. Spaldin, R. Ramesh, and M. Fiebig, Nat. Mater. **11**, 284 (2012).

[11] W. Wu, Y. Horibe, N. Lee, S.-W. Cheong, and J.R. Guest, Phys. Rev. Lett. **108**, 77203 (2012).

[12] C. Godau, T. Kämpfe, A. Thiessen, L.M. Eng, and A. Haußmann, ACS Nano **11**, 4816 (2017).

[13] E. Hassanpour, V. Wegmayr, J. Schaab, Z. Yan, E. Bourret, T. Lottermoser, M. Fiebig, and D. Meier, New J. Phys. **18**, 43015 (2016).

[14] J.R. Whyte and J.M. Gregg, Nat. Commun. **6**, 7361 (2015).

[15] P. Sharma, Q. Zhang, D. Sando, C.H. Lei, Y. Liu, J. Li, V. Nagarajan, and J. Seidel, Sci. Adv. **3**, 1 (2017).

[16] V. Janovec and J. Přívratská, in *Int. Tables Crystallogr.* (2006).

[17] E. Salje and H. Zhang, Phase Transitions **82**, 452 (2009).

[18] E.K.H. Salje and J.F. Scott, Appl. Phys. Lett. **105**, 252904 (2014).

[19] H. Yokota, H. Usami, R. Haumont, P. Hicher, J. Kaneshiro, E.K.H. Salje, and Y. Uesu, Phys. Rev. B **89**, 144109 (2014).

[20] S. Van Aert, S. Turner, R. Delville, D. Schryvers, G. Van Tendeloo, and E.K.H. Salje, Adv. Mater. **24**, 523 (2012).

[21] A.M. Glazer, Acta Crystallogr. Sect. B Struct. Crystallogr. Cryst. Chem. **28**, 3384 (1972).

[22] W.T. Lee, E.K.H. Salje, and U. Bismayer, Phase Transitions **76**, 81 (2003).

[23] L. Goncalves-Ferreira, S.A.T. Redfern, E. Artacho, and E.K.H. Salje, Phys. Rev. Lett. **101**, 97602 (2008).

[24] E.K.H. Salje, S. Li, M. Stengel, P. Gumbsch, and X. Ding, Phys. Rev. B **94**, 24114 (2016).

[25] M. Stengel, Phys. Rev. B **88**, 174106 (2013).

[26] E.A. Eliseev, A.N. Morozovska, Y. Gu, A.Y. Borisevich, L.-Q. Chen, V. Gopalan, and S. V. Kalinin, Phys. Rev. B **86**, 85416 (2012).

[27] R. Le Bihan, Ferroelectrics **97**, 19 (1989).

[28] N. Barrett, J.E. Rault, J.L. Wang, C. Mathieu, A. Locatelli, T.O. Mentes, M.A. Niño, S. Fusil, M. Bibes, A. Barthélémy, D. Sando, W. Ren, S. Prosandeev, L. Bellaiche, B. Vilquin, A. Petraru, I.P. Krug, and C.M. Schneider, J. Appl. Phys. **113**, 187217 (2013).

[29] J.E. Rault, W. Ren, S. Prosandeev, S. Lisenkov, D. Sando, S. Fusil, M. Bibes, A. Barthélémy, L.

Bellaiche, and N. Barrett, Phys. Rev. Lett. **109**, 267601 (2012).

[30] J.L. Wang, B. Vilquin, and N. Barrett, Appl. Phys. Lett. **101**, 92902 (2012).

[31] G.F. Nataf, P. Grysan, M. Guennou, J. Kreisel, D. Martinotti, C.L. Rountree, C. Mathieu, and N. Barrett, Sci. Rep. **6**, 33098 (2016).

[32] A. Migliori, J.L. Sarrao, W.M. Visscher, T.M. Bell, M. Lei, Z. Fisk, and R.G. Leisure, Phys. B Condens. Matter **183**, 1 (1993).

[33] D.J. Safarik, E.K.H. Salje, and J.C. Lashley, Appl. Phys. Lett. **97**, 111907 (2010).

[34] N.J. Perks, Z. Zhang, R.J. Harrison, and M.A. Carpenter, J. Phys. Condens. Matter **26**, 505402 (2014).

[35] R. Placeres-Jiménez, L.G. V Gonçalves, J.P. Rino, B. Fraygola, W.J. Nascimento, and J.A. Eiras, J. Phys. Condens. Matter **24**, 475401 (2012).

[36] J. Seidel, editor , *Topological Structures in Ferroic Materials* (Springer International Publishing, Cham, 2016).

[37] E.K.H. Salje, O. Aktas, M.A. Carpenter, V. V. Laguta, and J.F. Scott, Phys. Rev. Lett. **111**, 247603 (2013).

[38] O. Aktas, M.A. Carpenter, and E.K.H. Salje, Appl. Phys. Lett. **103**, 2011 (2013).

[39] O. Aktas, E.K.H. Salje, S. Crossley, G.I. Lampronti, R.W. Whatmore, N.D. Mathur, and M.A. Carpenter, Phys. Rev. B **88**, 174112 (2013).

[40] O. Aktas, S. Crossley, M.A. Carpenter, and E.K.H. Salje, Phys. Rev. B **90**, 165309 (2014).

[41] J. Sapriel, Phys. Rev. B **12**, 5128 (1975).

[42] S.A.T. Redfern, J. Phys. Condens. Matter **8**, 8267 (1996).

[43] M.A. Carpenter, A.I. Becerro, and F. Seifert, Am. Mineral. **86**, 348 (2001).

[44] See Supplemental Material for an XPS spectrum of the sample, a discussion about under and over-focusing conditions, MEM images after charge injection and maps of the surface potential.

[45] T. Zykova-Timan and E.K.H. Salje, Appl. Phys. Lett. **104**, 82907 (2014).

[46] J. Cazaux, J. Appl. Phys. **111**, 64903 (2012).

[47] B. Ziaja, R.A. London, and J. Hajdu, J. Appl. Phys. **99**, 33514 (2006).

[48] R. Ahluwalia, N. Ng, A. Schilling, R.G.P. McQuaid, D.M. Evans, J.M. Gregg, D.J. Srolovitz, and J.F. Scott, Phys. Rev. Lett. **111**, 165702 (2013).

[49] A. Biancoli, C.M. Fancher, J.L. Jones, and D. Damjanovic, Nat. Mater. **14**, 224 (2014).

[50] T. Tanaka, K. Matsunaga, Y. Ikuhara, and T. Yamamoto, Phys. Rev. B **68**, 205213 (2003).

[51] L. Wang, K.-J. Jin, J.-X. Gu, C. Ma, X. He, J. Zhang, C. Wang, Y. Feng, Q. Wan, J.-A. Shi, L. Gu, M. He, H.-B. Lu, and G.-Z. Yang, Sci. Rep. **4**, 6980 (2014).

# Supplemental material

## A. XPS ANALYSIS

Fig. S1 shows the Ti 2p XPS spectrum. It is used to monitor the concentration of oxygen vacancies[1]. A single, symmetric component is observed for the Ti $2p_{3/2}$ at 458.6 eV (FWHM: 0.65 eV) showing that all Ti is present in a fully oxidized $Ti^{4+}$ state. We can therefore exclude a significant presence of oxygen vacancies which would donate electrons, reducing $Ti^{4+}$ to $Ti^{3+}$, characterized by a low binding energy shoulder; and which could migrate and pin domain walls (DWs)[2].

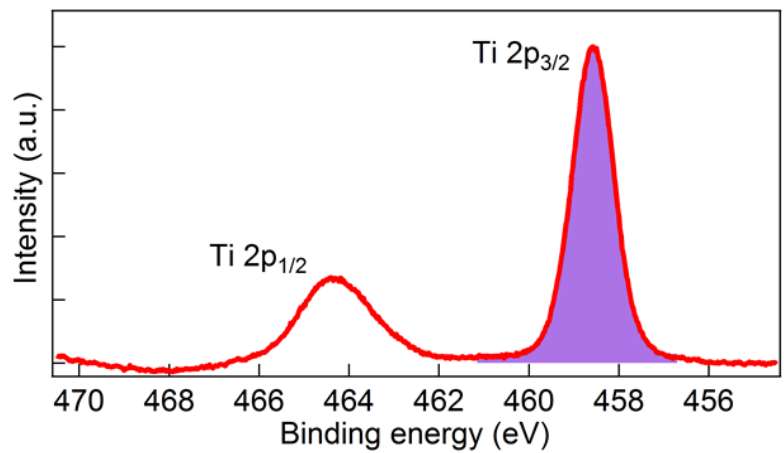

FIG. S1. XPS spectrum of the sample.

## B. PHYSICAL TOPOGRAPHY

In MEM, an asperity on the surface causes a local electric field perturbation that deviates incident and reflected electrons. The direction of the deviation depends on the geometry of the asperity: a dimple focuses the electrons (and acts like positive charge), giving a bright feature in direct imaging whereas a protrusion on the surface scatters and defocuses electrons[3]. The bright/dark contrast is inverted for an odd number of convergent lenses as in our microscope. This is why we observe bright dots on the images due to surface contamination.

## C. UNDER- AND OVERFOCUSING

In order to get a better understanding of the high contrast observed at $R_2$ and $V_2$, Fig. S2 shows an electron image obtained at SV = 0.3 V with a Field of View of 10 μm. The dots due to surface contamination are used to determine the best focus conditions. Two DWs are easily visible: $R_2$ and $V_2$. Previous studies have shown that the sign of a charged surface can be quickly identified by out-of-focus imaging[4,5]. Figure S2a and c are electron images obtained at SV = 0.3 V for under- and overfocusing conditions. Going from underfocus to overfocus conditions, the contrast at DWs goes from negative to positive values and the bright dots become dark.

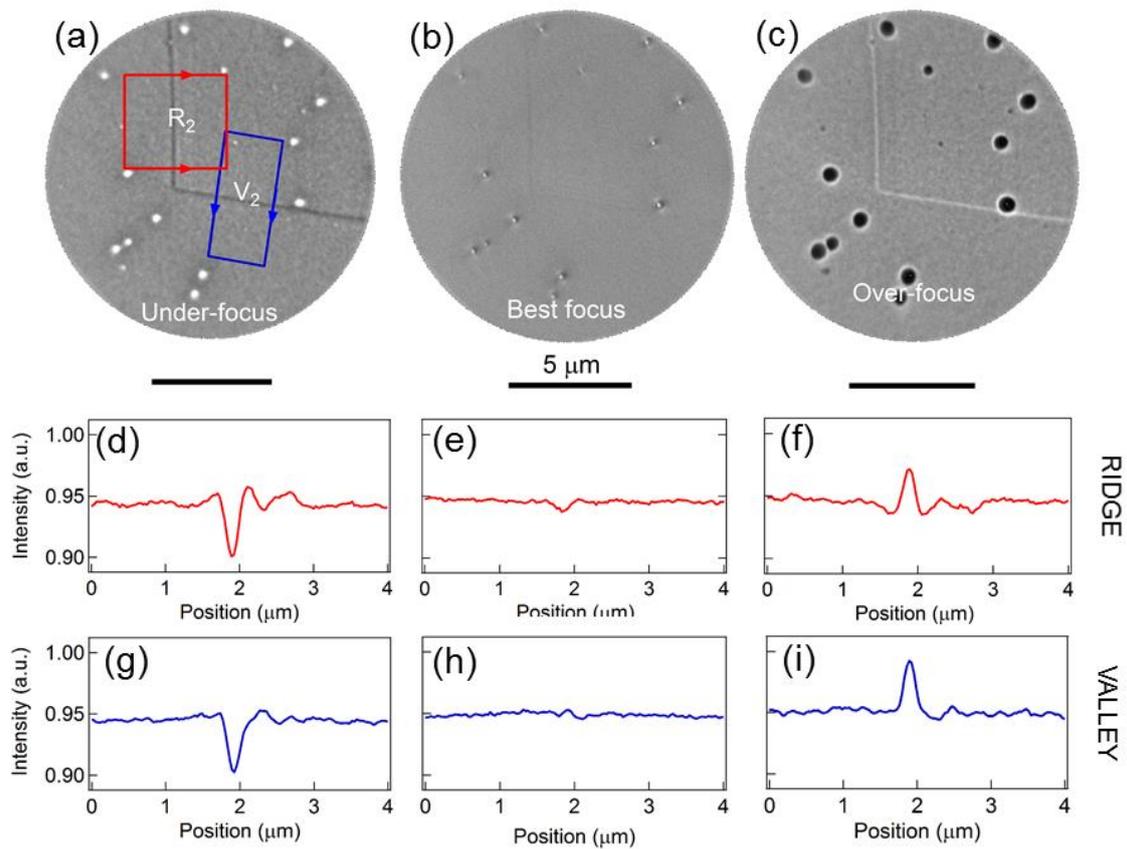

FIG. S2. MEM images of the surface. (a) at underfocus, (b) best focus and (c) overfocus. (d-f) Profiles extracted at $R_2$ for different focus conditions. (g-i) Profiles extracted at $V_2$ for different focus conditions.

The intensity profiles along DWs for different focus conditions are plotted in Fig.S2(d-i). At underfocus, the ridge $R_2$ appears as a thin dark line 400-500 nm wide, surrounded by two weaker bright lines (Fig. S1(a,d)). In the best focus conditions the DW is much fainter but still visible. It appears as a single dark line (Fig. S2(b)). Under-focus, $V_2$ also appears as a thin dark line extended on 500 nm, but surrounded by a single bright line on its right. In the best focus conditions $V_2$ is not visible. Overfocus, the contrast at DWs is the opposite of the contrast observed underfocus (Fig. S2(f,i)). The surface contamination shows the opposite behaviour, being bright underfocus and dark overfocus, as one would expect from purely physical topography.

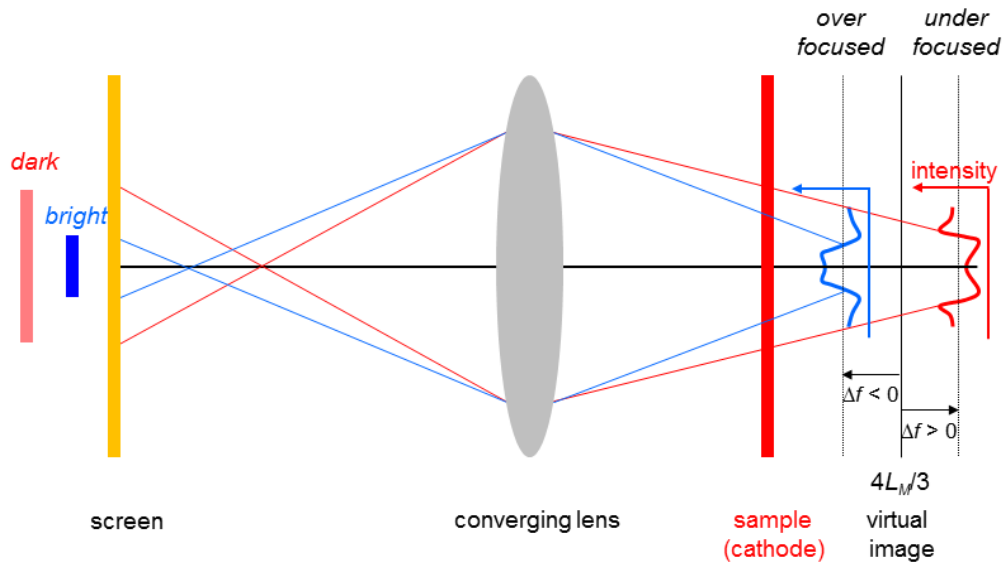

FIG. S3. Schematic showing how under- and overfocusing give rise to contrast in the microscope with an odd number of convergent lenses. The length of the vertical blue and pink lines, labelled bright and dark, on the left schematize the size of the image of the charged region and hence intensity in the case of an outward pointing polarity DW.

The shift in the MEM-LEEM transition with focus conditions is an indication that the surface potential variation at DWs is sharp. We suggest that the shift is due to sharp variations in electron turning distance from the surface: depending on the local surface potential, electrons are reflected at a different height above the surface. The rapidly varying local surface potential changes the z-motion of the electrons and therefore the image intensity in MEM as a function of SV. In particular, if the electron trajectories are strongly modified, electron rays cross and give rise to very bright features in MEM images known as caustics[6,7].

Figure S3 is a schematic showing how out-of-focus conditions determine the contrast between a polar domain wall and the neutral surface of the adjacent domains. The shift in the position of the virtual image behind the sample combines with the modulation of the surface potential near a domain wall of positive polarity to produce dark (bright) feature for under (over) focusing. The result is shown schematically by the virtual image intensity profiles.

## D. CHARGE INJECTION

In Fig. S4(a-c) we show images acquired at SV = 0.3 V after exposure to different energy electrons for 10 minutes. At higher SV more electrons are expected to penetrate into the sample, at the same time the inelastic mean free path decreases. Both effects increase the injected charge density. After irradiation at 6 V, the contrast at $V_1$, $V_2$, $V_3$ and $R_2$ is weaker but has not changed at $R_1$. After 10 V irradiation, the contrast at $R_2$ has disappeared, whereas the contrast at $R_1$ remains unchanged. We can also remark that the spots due to surface contamination are brighter, probably due to slight charging under the e-beam. Figure S4d shows the evolution of the number of electrons injected (blue curve) as a function of the start voltage. The number of electrons injected clearly increase above the MEM-LEEM transition at SV = 0.5 V. Then it decreases and finally saturates at a value of $3.5 \times 10^{16}$. We note that this is approximate since sample charging alters the reflectivity and the penetration depth varies rapidly with SV, however, it does provide an estimate for the effective total injected charge.

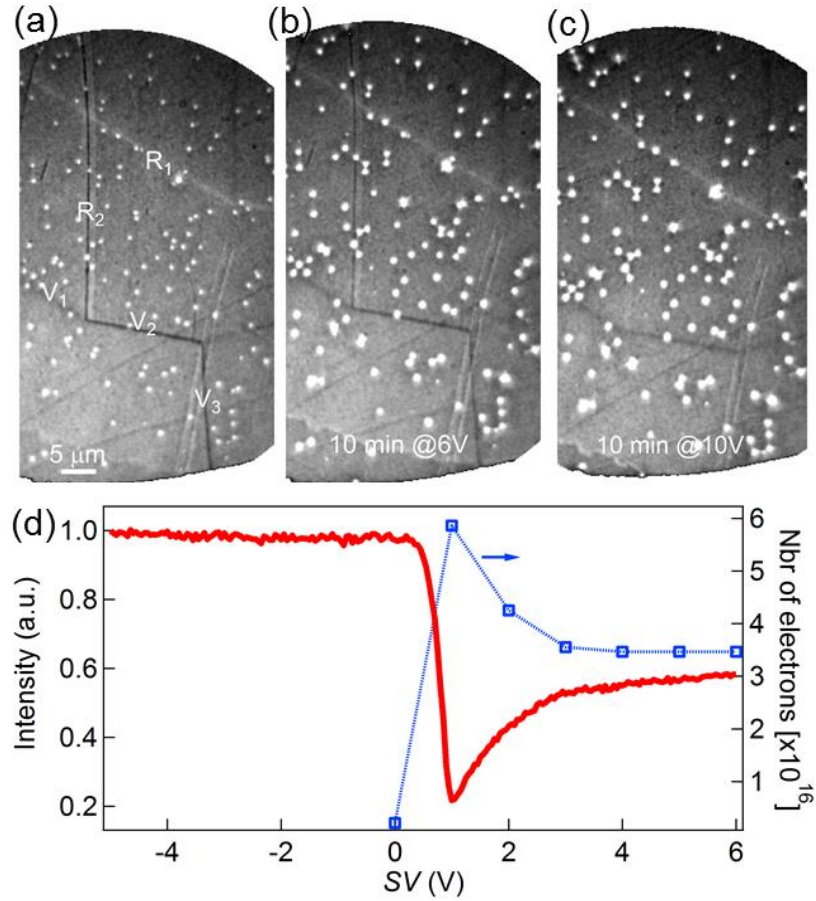

FIG. S4. Charge injection. (a) MEM images taken at SV = 0.3 V (a) before exposure to electrons, after exposure to electrons at (b) SV = 6 V and (c) SV = 10 V during 10 minutes. (d) MEM-LEEM curve in a domain and number of electrons injected in the beam spot during 10 minutes as a function of SV.

## E. SURFACE POTENTIAL

Electrons images acquired as a function of SV going from MEM to LEEM are used to determine the surface potential. At each point in the field of view the electron reflectivity curve is extracted from the image series. A pixel-by-pixel fit to the intensity using a complementary error function maps the MEM-LEEM transition and hence the local surface potential. The results are shown in Fig. S5(a,d). The surface potential is the same in domains on both sides of the DWs, as expected for the non-polar surface of $CaTiO_3$ whereas $R_2$ and $V_2$ show clear surface potential contrast with respect to the domains.

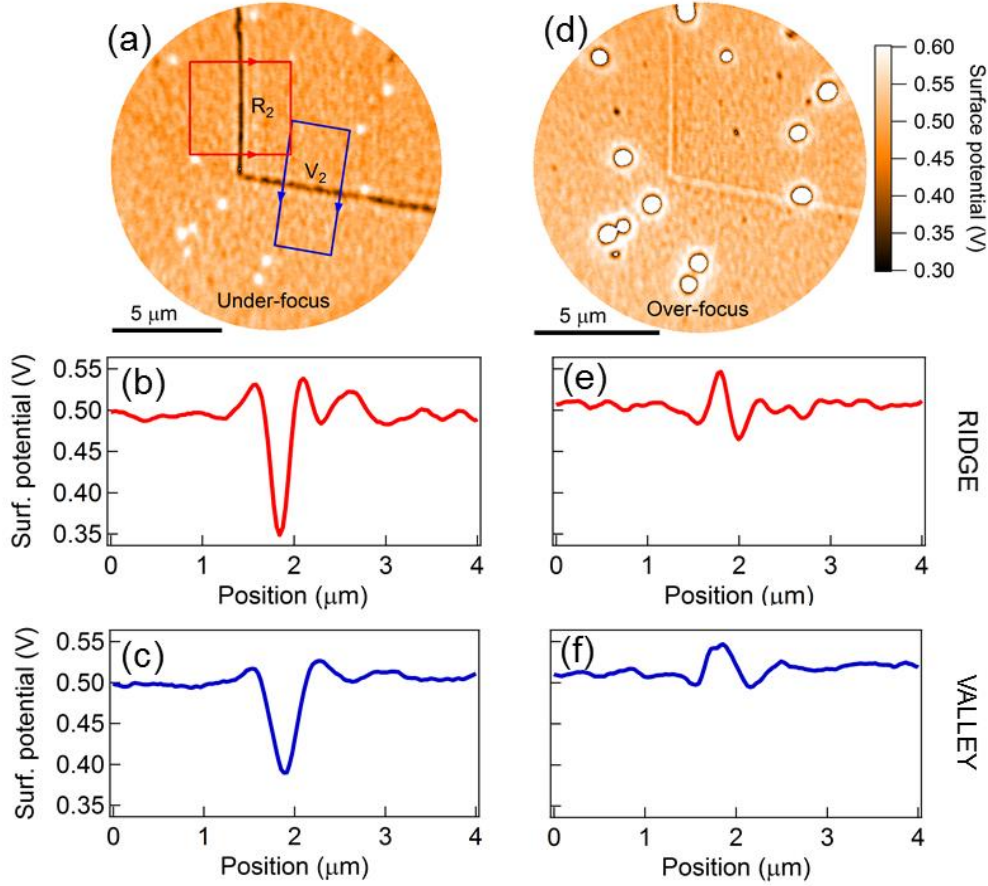

FIG. S5. For under-focus conditions: (a) surface potential map; profiles at (b) $R_2$ and (c) $V_2$. For over-focus conditions: (d) surface potential map; profiles at (e) $R_2$ and (f) $V_2$.

Figure S5(b) shows a profile across $R_2$ in underfocus conditions. The main feature has a surface potential lower by 140 mV with respect to the domains. In order to verify that the surface potential deduced from the MEM-LEEM curve can be used to determine the surface charge state, we observe the effect on the DW contrast of overfocusing conditions. As shown in Fig. S5(a,d), surface potential inversion occurs at DWs when changing from under- to overfocusing conditions. This is confirmed by the profiles in Fig. S5(b,c) and S5(e,f) respectively.

# REFERENCES

[1] J.E. Rault, J. Dionot, C. Mathieu, V. Feyer, C.M. Schneider, G. Geneste, and N. Barrett, Phys. Rev. Lett. **111**, 127602 (2013).

[2] T. Tanaka, K. Matsunaga, Y. Ikuhara, and T. Yamamoto, Phys. Rev. B **68**, 205213 (2003).

[3] S.A. Nepijko and N.N. Sedov, in *Adv. Imaging Electron Phys.* (1997), pp. 273–323.

[4] S.M. Kennedy, C.X. Zheng, W.X. Tang, D.M. Paganin, and D.E. Jesson, Proc. R. Soc. A Math. Phys. Eng. Sci. **466**, 2857 (2010).

[5] C.L. Sones, S. Mailis, W.S. Brocklesby, R.W. Eason, and J.R. Owen, J. Mater. Chem. **12**, 295 (2002).

[6] S.M. Kennedy, C.X. Zheng, W.X. Tang, D.M. Paganin, and D.E. Jesson, Ultramicroscopy **111**, 356 (2011).

[7] S.M. Kennedy, D.E. Jesson, and D.M. Paganin, IBM J. Res. Dev. **55**, 3:1 (2011).